\documentstyle[11pt,newpasp,twoside,epsf]{article}
\def\edcomment#1{\iffalse\marginpar{\raggedright\sl#1\/}\else\relax\fi}
\reversemarginpar

\begin{document}
\title{A Time Dependent Model of HD209458b}
 \author{N. Iro, B. B\'ezard}
\affil{Observatoire de Paris-Meudon, LESIA,\\
 5 place J. Janssen, 92195 Meudon cedex, France}
\author{T. Guillot}
\affil{Observatoire de la C\^{o}te d'Azur, Laboratoire CASSINI,\\
BP 4229, 06304 Nice Cedex 4, France}

\begin{abstract}
We developed a time-dependent radiative model for 
the atmosphere of HD209458b
to investigate its thermal structure and chemical composition.
Time-dependent temperature profiles were calculated, using 
a uniform zonal wind modelled as a solid body rotation.
We predict day/night temperature variations of 600K around 0.1 bar,
for a 1 km/s wind velocity, 
in good agreement with the predictions by Showman \& Guillot (2002).
On the night side, the low temperature allows the sodium to condense.
Depletion of sodium in the morning limb may explain the lower than expected abundance found by Charbonneau et al (2002).
\end{abstract}

\section{Introduction}
The transiting planet HD209458b should be in synchronous rotation.
This implies a significant day/night temperature variation driving a strong atmospheric circulation
with wind velocity of $\sim$1 km/s
(Guillot \& Showman, 2002).
Charbonneau et al. (2002) constrained the sodium abundance in its atmosphere by a spectroscopic study.
It is lower than predicted by most models
which assume a uniform temperature field.
We present a time-dependent radiative model for this atmosphere.
It allows us, by modeling the circulation as a solid body rotation, to investigate the longitudinal dependence of temperature.

\section{Time-dependent radiative model}
Radiative fluxes in the atmosphere (stellar and thermal infrared) are calculated throught a line by line code in the spectral range 0.3--9 $\mu$m.
The following opacity sources were included: Rayleigh scattering; H$_2$ and He CIA absorptions;
molecular species such as H$_2$O, CO, CH$_4$, Na, K, TiO; H$^-$ bound-free and H$_2^-$ free-free absorptions.
Clouds were not considered.
We calculated molecular abundances assuming thermochemical equilibrium at each level and a solar abundance of the elements.
Our bottom boundary condition is set by the intrinsic flux of the planet.
Guillot \& Showman (2002) showed that an equivalent intrinsic effective temperature of T$_{\rm int}$=400K is required to fit HD209458b's age and radius.

The solution profile is calculated by solving the energy equation:\\
$\frac{{\rm d}T}{{\rm d}t}=-\frac{m \ g}{c_p}\left(\left(-\frac{{\rm d}F_\star}{{\rm d}P}\right)-\left(-\frac{{\rm d}F_{IR}}{{\rm d}P}\right)\right)
{\rm , }$
where $-\frac{{\rm d}}{{\rm d}P}F_{\star/IR}$
are respectively the heating/cooling rates.
A time-marching algorithm is used to reach a periodic solution.
We investigated the temperature as a function of longitude by modelling the atmospheric circulation as a solid body rotation.
Fig. 1 shows the results for a 1 km/s wind velocity, as
predicted by Showman \& Guillot (2002).
We found a 600K day/night temperature variationat the 0.1-bar level, in agreement with their model.
A major effect is the condensation of Na on the night side.
The sodium absorption present in the HST observations is a factor of 3 less than predicted by most static models (Charbonneau et al., 2002).
If the sodium condenses in the night side, half of the limb is depleted in sodium, which might explain the weak HST absorption.

\begin{figure}  
\begin{center}
\epsfclipon
\epsfxsize=10.5cm
\epsffile{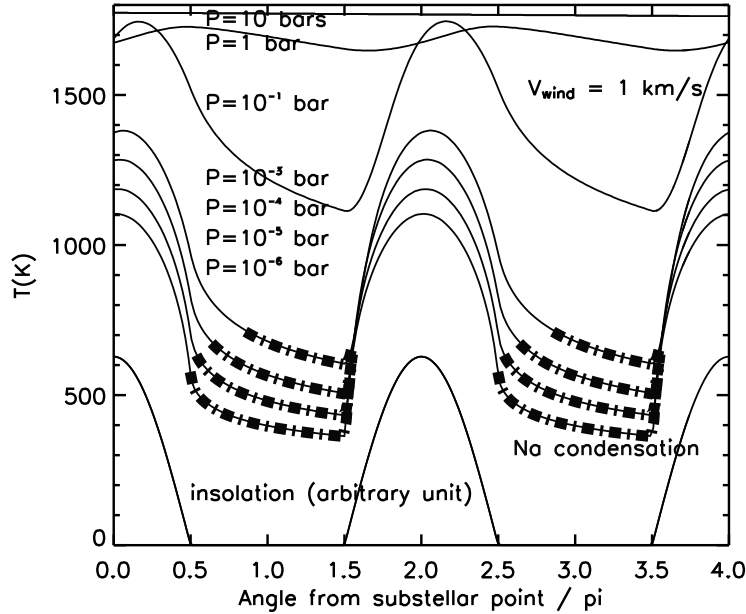}
\leavevmode
\caption{Temperature at some levels as a function of longitude.}
\end{center}
\end{figure}

\section{References}
Charbonneau, D., Brown, T.~M., Noyes, 
R.~W., \& Gilliland, R.~L. 2002, \apj, 568, 377 \\
Guillot, T. \& Showman, A.~P., 2002, \aap, 385,156\\
Showman, A.~P. \& Guillot, T., 2002, \aap, 385,166
\end{document}